\documentclass[preprint,showpacs,preprintnumbers,amsmath,amssymb]{revtex4}

\usepackage{graphicx}      % needed for figures
\usepackage{dcolumn}      % needed for some tables
\usepackage{bm}             % for math
\usepackage{amssymb}    % for math

\begin{document}

\title{Appearance of vortices and monopoles in a decomposition of an SU(2) Yang-Mills field}
\author{A. Mohamadnejad} \altaffiliation {a.mohamadnejad@ut.ac.ir}
\author{S. Deldar} \altaffiliation {sdeldar@ut.ac.ir}
\affiliation{Department of Physics, University of Tehran, P.O. Box 14395/547, Tehran 1439955961, Iran}

\begin{abstract}
A decomposition for the SU(2) Yang-Mills field in the low-energy limit is obtained by supposing that, in
the low-energy limit, the field strength tensor of an SU(2) Yang-Mills field can be obtained by 
multiplying two parts, $ G_{\mu\nu} $ and $ \textbf{n} $, such that $ \textbf{G}_{\mu\nu}=G_{\mu\nu} \textbf{n} $.  $ G_{\mu\nu} $ 
is a space-time tensor and $ \textbf{n} $ is an isotriplet unit vector field that gives the Abelian direction at each
space-time point.
By  Abelianizing the field strength tensor $ \textbf{G}_{\mu\nu} $, we show how (singular) vortices
and monopoles can appear in the low-energy limit of SU(2) Yang-Mills theory.
If  the decomposition is valid on the  boundary of the system, then we show that vortices with finite string tension can also appear.
The interesting point is that we have started with a decomposed Yang-Mills
 field and we have ended up with a theory with an Abelian gauge field coupled to a scalar field.  
The effect of this scalar field on monopoles is also discussed.
\end{abstract}


\maketitle

\section{\label{sec:level1}Introduction}

It is stated that ultraviolet and infrared limits of a Yang-Mills theory characterize different regimes.
The qualitative picture of this assertion is built especially by 't Hooft and Polyakov \cite{'Hooft}.
In the ultraviolet limit, the Yang-Mills theory is asymptotically free and perturbative methods are adequate.
This limit describes the interaction between massless gluons which correspond to the transverse polarizations
of the gauge field $ A_{\mu} $. At low energies the Yang-Mills theory becomes strongly coupled and
perturbative techniques fail, so nonperturbative methods must be developed. On the other hand,
Yang-Mills theories in the low-energy limit must exhibit color confinement. Therefore, describing the
confinement problem needs nonperturbative methods and it has long been argued that the confinement can take
place through the condensation of monopoles which leads to the dual Meissner effect in a dual superconductor \cite{Nambu}.
For a review of the dual superconductor picture, see \cite{Ripka}.

Nonperturbative effects in the low-energy limit can be presented effectively by the topological structures
of the gauge theory such as vortices and monopoles. The underlying gauge symmetry can be  represented by the nontrivial topological degrees
of freedom. There are some methods for extracting these topological
degrees of freedom in the pure Yang-Mills theory, e.g., Abelian projection \cite{Abelian}
and field decomposition \cite{Cho,Faddeev,Shabanov}. Abelian projection is a partial gauge fixing in which
the projected gauge fields contain singularities interpreted as topological defects. In the second method,
Abelian decomposition, one can do the same thing without gauge fixing \cite{Shabanov}. The theories that resul
from these decompositions consist of a well defined and self-consistent subset of a non-Abelian gauge theory
for a given symmetry group. These theories are restricted and the dynamical degrees of freedom are reduced, 
providing us with a self-consistent but nontrivial subset of the original gauge theory. As a result of 
these methods, the original Yang-Mills theory turns into electrodynamics with magnetic monopoles, and they lead to
Abelian dominance \cite{Yotsuyanagi} and magnetic monopole dominance \cite{Stack}. Indeed,
in both the Abelian projection \cite{Suganuma} and field decomposition approaches \cite{Deldar}, one
can obtain Wu-Yang magnetic monopoles \cite{Wu}, and based on their condensations the potential of
a static quark-antiquark pair is derived. In agreement with lattice results, this potential is composed of two parts: the
first part is a Yukawa term that dominates the ultraviolet limit or the short-range limit, and the second part
is a linear term responsible for the confinement that dominates the low-energy limit which is valid at large distances.

Even though the gauge field $ A_{\mu} $ is a
proper order parameter for describing the theory in its high-energy limit, in the low-energy limit
some other parameters become more appropriate. Therefore, one can decompose
the Yang-Mills field to new collective variables that are more appropriate for describing the low-energy limit. Decomposing the Yang-Mills fields
has been done by various methods. These decompositions pursue different purposes, in particular, in
connection with the issue of quark confinement in quantum chromodynamics (QCD). In Cho's restricted SU(2) Yang-Mills theory \cite{Cho},
there are four degrees of freedom: two  dynamical and two topological while an SU(2) Yang-Mills
field has six dynamical degrees of freedom. Hence, it seems that in the infrared limit, some degrees of freedom
do not play considerable roles.
One can extend Cho's restricted theory so that it consists all six dynamical degrees of freedom of an SU(2) Yang-Mills field  \cite{Cho}. A 
unified treatment of both monopoles and center
vortices within the scenario of Cho decomposition can be found in Ref. \cite{Oxman}.
In the Faddeev-Niemi method , a special form of Cho decomposition is studied \cite{Faddeev}.
Faddeev and Niemi declare that their decomposition of SU(2) Yang-Mills theory is complete, but this assertion has been criticized recently and their 
reformulation is inequivalent to Yang-Mills theory \cite{Evslin,Niemi3}.
However, One can obtain Skyrme-Faddeev
Lagrangian by integrating out some of the new variables appearing in their decomposition \cite{Pak}. The Skyrme-Faddeev theory, like QCD, is 
a theory of confinement that confines the magnetic flux of the monopoles \cite{Bae}. In addition, Faddeev and Niemi have investigated the possibility that the
low-energy spectrum of pure Yang-Mills theory can be constructed of closed and knotted strings as stable solitons \cite{Faddeev}.
They derive an off-shell generalization of their decomposition \cite{Niemi1} and they also present a new
decomposition that realizes explicitly a symmetry between electric and magnetic variables, suggesting a
duality picture between the corresponding phases \cite{Niemi2}.

In this paper, we propose a decomposition in the low-energy limit that has the same field
strength tensor form as Cho's restricted theory, $ \textbf{G}_{\mu\nu}=G_{\mu\nu} \textbf{n} $, however, the form of the decomposed Yang-Mills field is different from Cho's 
restricted field.
It seems that this decomposition describes topological structures including both vortices and monopoles 
which dominate the nonperturbative regime of the theory.
We show that this decomposition for the low-energy limit supports vortices as topological solitons. Topology provides the existence arguments.
In addition to vortices, Wu-Yang monopoles 
can also be obtained in this decomposition.
However, there are great contrasts between these monopoles and vortices: unlike vortices, Wu-Yang monopoles are not topological solitons
and they do not have finite energy. Finally, we discuss how the scalar field that appears in our decomposition can affect the magnetic field.
In Paticular, monopoles cannot appear in the Higgs vacuum of the system. In other words, magnetic monopoles are confined.

In the following sections, we introduce our decomposition in the low-energy limit by defining $ \textbf{G}_{\mu\nu} $ such that 
$ \textbf{G}_{\mu\nu}=G_{\mu\nu} \textbf{n} $.
We propose a decomposition for the Abelian gauge field. By some topological arguments, we show that a vortex solution is possible in principle. 
We also show that
the scalar field that appears in our decomposition provides a medium that affects the magnetic field and makes it zero in the Higgs vacuum of the system. 
Therefore, monopoles are somehow confined.
Finally, the summary and discussions are given.

\section{\label{sec:level2}SU(2) Yang-Mills field in the low-energy limit}

\begin{figure}[]
\centering
\includegraphics*[width=9cm]{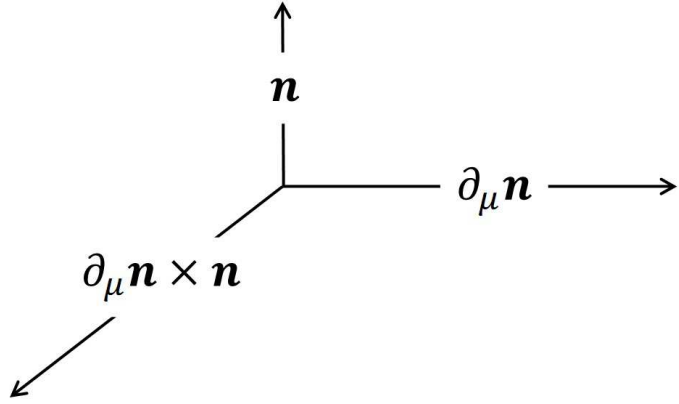}
\caption{Constructing an orthogonal basis for the SU(2) color space at each space-time point.} \label{Fig. (1)}
\end{figure}

In this paper, to avoid unnecessary complications, we concentrate on the SU(2) gauge group which is the simplest non-Abelian group.
Motivated by Cho's restricted theory, we
assume that  within a good approximation, the following form of the field strength tensor $ \textbf{G}_{\mu\nu} $, is dominant
in the low-energy limit of the SU(2) Yang-Mills theory
\begin{equation}
\textbf{G}_{\mu\nu}=G_{\mu\nu} \textbf{n}. \label{1}
\end{equation}
$ G_{\mu\nu} $ is a colorless tensor and $ \textbf{n} $ is a three-components unit vector field pointing in the color direction. We make a decomposition by constructing  an orthogonal basis
for the color space by $ \textbf{n} $ and its derivatives, e.g., $ \textbf{n} $, $ \partial_{\mu} \textbf{n} $,
and $ \textbf{n}\times\partial_{\mu} \textbf{n} $ shown in Fig. (\ref{Fig. (1)})
$$ \textbf{n} . \partial_{\mu} \textbf{n} = \textbf{n} . (\textbf{n}\times\partial_{\mu} \textbf{n})
= \partial_{\mu} \textbf{n} . (\textbf{n}\times\partial_{\mu} \textbf{n}) = 0 . $$
Then we expand the SU(2) Yang-Mills field as follows:
\begin{equation}
\textbf{A}_{\mu}=C_{\mu} \textbf{n} + \phi_{1} \partial_{\mu} \textbf{n} + \phi_{2}  \textbf{n}\times\partial_{\mu} \textbf{n}, \label{2}
\end{equation}
where $ C_{\mu} $, $ \phi_{1} $, and $ \phi_{2} $ are the coefficients of the expansion.

Because of Eq. (\ref{1}), $ C_{\mu} $, $ \phi_{1} $, and $ \phi_{2} $ are not independent. In the following, we find the relation between them.
For the SU(2) field strength tensor, we have
\begin{equation}
\textbf{G}_{\mu\nu} = \partial_{\mu} \textbf{A}_{\nu} - \partial_{\nu} \textbf{A}_{\mu} + g \textbf{A}_{\mu} \times \textbf{A}_{\nu}. \label{3}
\end{equation}
Substituting Eq. (\ref{2}) into Eq. (\ref{3}) and requiring $ \textbf{G}_{\mu\nu} $ to have the form of Eq. (\ref{1}),
these relations between $ \phi_{1} $, $ \phi_{2} $, and $ C_{\mu} $ are obtained:
\begin{eqnarray}
\partial_{\mu} \phi_{1} - C_{\mu} (1 + g \phi_{2}) = 0, \nonumber\\ &&
\hspace{-41mm} \partial_{\mu} \phi_{2} - g C_{\mu}  \phi_{1} = 0. \label{4}
\end{eqnarray}
Changing the variables
\begin{eqnarray}
\phi_{1} = \frac{\rho}{g^{2}}, \nonumber\\ &&
\hspace{-21mm} 1 + g \phi_{2} = \frac{\sigma}{g}, \label{5}
\end{eqnarray}
where $ \rho $ and $ \sigma $ are real scalar fields and applying these new variables to Eq. (\ref{4}), one obtains
\begin{eqnarray}
\partial_{\mu} \rho - g C_{\mu} \sigma = 0, \nonumber\\ &&
\hspace{-31mm} \partial_{\mu} \sigma + g C_{\mu} \rho = 0. \label{6}
\end{eqnarray}
The above equations can be written in a compact form
\begin{eqnarray}
D_{\mu} \phi = 0, \nonumber \label{7}
\end{eqnarray}
where $ D_{\mu} $ is the covariant derivative, $ D_{\mu} = \partial_{\mu} + i g C_{\mu} $, and $ \phi $
is a complex scalar field, $ \phi = \rho + i \sigma $. Notice that the coupling between the scalar field $ \phi $,
and the Abelian gauge field $ C_{\mu} $, is the same as the coupling of the original SU(2) gauge theory, $ g $.
Therefore, we conclude that the following decomposition for the SU(2) Yang-Mills field satisfies Eq. (\ref{1})
\begin{equation}
\textbf{A}_{\mu}=C_{\mu} \textbf{n} + \frac{1}{g} \partial_{\mu} \textbf{n} \times \textbf{n}
+ \frac{\rho}{g^{2}} \partial_{\mu} \textbf{n} + \frac{\sigma}{g^{2}} \textbf{n}\times\partial_{\mu} \textbf{n}, \label{8}
\end{equation}
with a constraint
\begin{equation}
D_{\mu} \phi  = (\partial_{\mu} + i g C_{\mu})(\rho + i \sigma) = D_{\mu} \rho  + i D_{\mu} \sigma = 0, \label{9}
\end{equation}
where $ D_{\mu} \rho = \partial_{\mu} \rho - g C_{\mu} \sigma $ and $ D_{\mu} \sigma = \partial_{\mu} \sigma + g C_{\mu} \rho $. 
The above condition must be satisfied in the infrared regime of the theory to get Eq. (\ref{1}).
Note that we have started with a Yang-Mills field $\textbf{A}_{\mu}$ and by this decomposition,  we have obtained an Abelian gauge field $C_{\mu}$
and a scalar field $\phi$ coupled to it.  In Sect. \ref{sec:level4}, we show that the constraint on the field $\phi$  via Eq.  (\ref{9}) which relates these two fields to each other, leads 
to the appearance of vortices in the theory.

Equation (\ref{9}) shows how the fields  $ \rho $, $ \sigma $, and $ C_{\mu} $ depend on each other in the infrared regime.
A trivial solution for Eq. (\ref{9}) is
\begin{eqnarray}
\rho = \sigma = 0, \label{10}
\end{eqnarray}
which leads to Cho's restricted theory with four degrees of freedom,
two topological degrees of freedom for $ \textbf{n} $ and two dynamical degrees of freedom for $ C_{\mu} $ corresponding to two polarizations. The field strength tensor for Cho's restricted theory is
\begin{equation}
\textbf{G}_{\mu\nu}=\lbrace \partial_{\mu}  C_{\nu} - \partial_{\nu}  C_{\mu}  - \frac{1}{g} \textbf{n} .  (\partial_{\mu} \textbf{n} \times \partial_{\nu} \textbf{n}) \rbrace \textbf{n} \label{11}
\end{equation}
Notice that Eq. (\ref{9}) is familiar, indeed Cho found the restricted SU(2) Yang-Mills field by a similar condition
\begin{equation}
\triangledown_{\mu} \textbf{n} = (\partial_{\mu} + g \textbf{A}_{\mu} \times) \textbf{n} = 0 \,  \Rightarrow
\, \textbf{A}_{\mu}=C_{\mu} \textbf{n} + \frac{1}{g} \partial_{\mu} \textbf{n} \times \textbf{n}, \label{12}
\end{equation}
where $ C_{\mu} = \textbf{A}_{\mu}.\textbf{n} $.
Hence, it seems that in the infrared limit of an Abelian or non-Abelian Yang-Mills theory,
we have a Yang-Mills field that obliges the covariant derivative of scalar fields to be zero, $ \triangledown_{\mu} \textbf{n} = 0$.

One can overlook the condition (\ref{9}) to generalize Eq. (\ref{8}). Then Eq. (\ref{1}) is no longer valid. The result is Faddeev-Niemi decomposition \cite{Faddeev} which leads to the following field strength tensor
\begin{eqnarray}
\textbf{G}_{\mu\nu}=\lbrace F_{\mu\nu} +  (1-\frac{\rho^{2} + \sigma^{2} }{g^{2}}) H_{\mu\nu} \rbrace \textbf{n}  \nonumber\\ &&
\hspace{-58mm} +\frac{1}{g^{2}} (D_{\mu} \rho \partial_{\nu} \textbf{n} - D_{\nu} \rho \partial_{\mu} \textbf{n} )  \nonumber\\ &&
\hspace{-58mm} +\frac{1}{g^{2}} (D_{\mu} \sigma \textbf{n} \times \partial_{\nu} \textbf{n} - D_{\nu} \sigma \textbf{n} \times \partial_{\mu} \textbf{n} ) , \label{99}
\end{eqnarray}
where
\begin{eqnarray}
F_{\mu\nu} = \partial_{\mu}  C_{\nu} - \partial_{\nu}  C_{\mu} , \nonumber\\ &&
\hspace{-40mm} H_{\mu\nu} = - \frac{1}{g} \textbf{n} .  (\partial_{\mu} \textbf{n} \times \partial_{\nu} \textbf{n}) .
\end{eqnarray}

In the next section, we find a nontrivial solution for Eq. (\ref{9}) and propose a decomposition
for the U(1) gauge field $ C_{\mu} $.

\section{\label{sec:level3} Abelian gauge field decomposition}

The constraint, Eq. (\ref{9}), which is part of our decomposition is by itself quite strong and worth
studying independently. We show how this constraint restricts
the Abelian U(1) gauge field and leads to the appearance of string-like (vortex) objects.
It is similar to the case where the condition of Eq. (\ref{12}) leads to the appearance of monopoles \cite{Cho}. 
Equation (\ref{9}) implies
\begin{equation}
\partial_{\mu} (\rho^{2} + \sigma^{2}) = 0 \quad \Rightarrow \quad  \phi^{\ast} \phi= \rho^{2} + \sigma^{2} = a^{2} , \label{13}
\end{equation}
where $ a $ is a constant. Notice that in Cho's restricted theory $ a $ is zero, but here it is non-zero. The non-zero value of $ a $ leads to some interesting differences between the decomposition 
here and the Cho's original decomposition, and plays an essential role in the appearance of vortices.
Equation (\ref{9}) can be solved exactly for $ C_{\mu} $
\begin{eqnarray}
C_{\mu} = \frac{1}{g a^{2}} (\sigma \partial_{\mu} \rho - \rho \partial_{\mu} \sigma). \label{14}
\end{eqnarray}
In the above equation, the Abelian gauge field  $C_{\mu}$ is decomposed to the scalar fields $\sigma$ and $\rho$.

The field strength tensor can be written in terms of electric and magnetic field strength tensors, $F_{\mu\nu}$ and $ B_{\mu\nu} $, respectively
\begin{equation}
\textbf{G}_{\mu\nu} = (F_{\mu\nu} + B_{\mu\nu}) \textbf{n} , \label{15}
\end{equation}
where
\begin{eqnarray}
F_{\mu\nu} = \partial_{\mu} C_{\nu} - \partial_{\nu} C_{\mu}, \nonumber\\ &&
\hspace{-38mm} B_{\mu\nu} = -\frac{1}{g^{\prime}} \textbf{n} .  (\partial_{\mu} \textbf{n} \times \partial_{\nu} \textbf{n}) = -\frac{1}{g} (1-\frac{\rho^{2} + \sigma^{2} }{g^{2}}) \textbf{n} .  (\partial_{\mu} \textbf{n} \times \partial_{\nu} \textbf{n}). \label{16}
\end{eqnarray}
Notice that since $ \rho^{2} + \sigma^{2} = a^{2} $, the contributions of $ \rho $ and $ \sigma $ are included in the new coupling $ g^{\prime} $ , where
\begin{equation}
\frac{1}{g^{\prime}} =  \frac{1}{g} - \frac{a^{2}}{g^{3}}. \label{17}
\end{equation}
We take $ a \leqslant g $ in order to have $ g^{\prime} \geqslant 0 $. Note that the coupling constant increases in the low-energy limit,
 $ g^{\prime} \geqslant g $, which is in agreement with the behavior of the coupling constant in Yang-Mills theories.

In the next section, we use Eqs. (\ref{13}) and  (\ref{14}) to obtain (singular) vortices. However, if  these equations are valid only on the  boundary of the system, then we get vortices with finite string tensions.

\section{\label{sec:level4}Vortices; an existence argument}

Vortices, which are classical string-like objects, appear in this theory as topological objects.
To observe vortices, we take the boundary of the space to be a circle at infinity,
denoted by $ S^{1}_{R} $.
Vortices are described by the homotopy class of a mapping $ \Pi_{1} (S^{1}) $
of the spatial circle $ S^{1}_{R} $ to the coset space $ S = U(1) $ of the internal space.
Now to define this mapping, one needs a two-components scalar field in the theory,
at least on $ S^{1}_{R} $. The scalar fields $ \rho $ and $ \sigma $ in decomposition (\ref{14}) can be used to define the
mapping $ \Pi_{1} (S^{1}) $. We define the topological charge by
the homotopy class of the mapping $ \Pi_{1} (S^{1}) $ given by $ (\rho,\sigma) $
\begin{equation}
(\rho,\sigma); \quad S^{1}_{R} \rightarrow S^{1} = U(1). \label{18}
\end{equation}

\begin{figure}[tbp]
\centering
\includegraphics*[width=7cm]{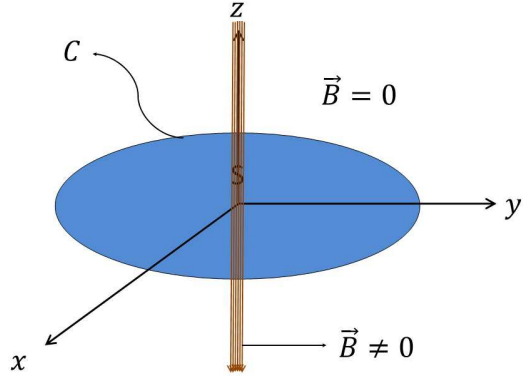}\hspace{1cm}
\caption{The magnetic field is zero everywhere but not on the z axis} \label{Fig. (2)}
\end{figure}

With this opening comment we show how to extract vortices.
The homotopy class $ \Pi_{1} (S^{1}) $ defined by
the following ansatz describes the vortex with a unit flux tube
\begin{equation}
(\rho,\sigma)=a\frac{\overrightarrow{r}}{r}=a (cos(\varphi),sin(\varphi)) , \label{19}
\end{equation}
where $ \varphi $ is the azimuthal circular coordinate of $ S^{1}_{R} $ and $ r $ is the distance from the $ z $ axis in the cylindrical coordinate system.
Using Eq. (\ref{19}) in Eq. (\ref{14}) one obtains
\begin{eqnarray}
C_{\mu} = - \frac{1}{g} \partial_{\mu} \varphi,  \nonumber\\ &&
\hspace{-30mm} \Longrightarrow C_{r} = C_{z} = 0 \, , \, \,\,\,\, C_{\varphi} = - \frac{1}{gr} . \label{20}
\end{eqnarray}
The magnetic field $ \overrightarrow{B} $ is obtained as the following
\begin{eqnarray}
\overrightarrow{B} = \overrightarrow{\nabla} \times \overrightarrow{C} =
\widehat{r} (\frac{1}{r} \frac{\partial C_{z}}{\partial \varphi} -  \frac{\partial C_{\varphi}}{\partial z}) +
\widehat{\varphi} ( \frac{\partial C_{r}}{\partial z} -  \frac{\partial C_{z}}{\partial r} )  \nonumber\\ &&
\hspace{-62mm} + \widehat{k} \frac{1}{r} ( \frac{\partial (r  C_{\varphi})}{\partial r} -  \frac{\partial C_{r}}{\partial \varphi} ) = 0  . \label{21}
\end{eqnarray}
Indeed, in general we have
\begin{eqnarray}
(\rho,\sigma)=a (cos(\alpha),sin(\alpha)) \Rightarrow C_{\mu} = - \frac{1}{g} \partial_{\mu} \alpha
\Rightarrow F_{\mu\nu} = 0 .
\end{eqnarray}
The above calculations are true  every place in space, but not on the z axis where $ r=0 $.
The magnetic flux passing through the closed curve C in Fig. (\ref{Fig. (2)}), is not zero:
\begin{eqnarray}
\phi_{B} = \int_{S} \overrightarrow{B} .
 \overrightarrow{ds} = \int_{S} (\overrightarrow{\nabla} \times \overrightarrow{C}) .
 \overrightarrow{ds} = \oint_{C} \overrightarrow{C} . \overrightarrow{dl}  \nonumber\\ &&
\hspace{-64mm} = \int^{2\pi}_{0}  - \frac{1}{gr} rd\varphi = - \frac{2\pi}{g}. \label{221}
\end{eqnarray}

It shows that on the z axis the magnetic field is singular as well as $ C_{\varphi} $ in Eq. (\ref{20}).
So, although the magnetic field is zero everywhere, there exists an infinite magnetic field on the z axis, responsible for the magnetic flux of Eq. (\ref{221}),
which is evidence of a string-like object (vortex) lying on the z axis. One can obtain the magnetic field from Eq. (\ref{221})
\begin{eqnarray}
\int_{S} \overrightarrow{B} . \overrightarrow{ds} = \int^{R}_{0} \int^{2\pi}_{0} B \, rd\theta \, dr =  - \frac{2\pi}{g} \nonumber\\ &&
\hspace{-55mm} \Longrightarrow \int^{R}_{0} B \, rdr  = -\frac{1}{g} \nonumber\\ &&
\hspace{-55mm} \Longrightarrow B = -2 \frac{\delta(r)}{gr}, \label{231}
\end{eqnarray}
therefore, the string 
tension of such a vortex is infinite \cite{Mo}.

To get vortices with finite string tension, we suggest that Eq. (\ref{13}) and (\ref{14}), are valid just for the boundary of the vortex solution, $ r \rightarrow \infty $, not everywhere. Indeed, we suppose that the decomposition (\ref{8}) with the constraint (\ref{9}) is true for the low-energy limit or the boundary of the system, and for the vortex core, the decomposition (\ref{8}) is still true without the constraint (\ref{9}).
Therefore, if we consider Eqs. (\ref{19}), eq. (\ref{20}) and eq. (\ref{21}) as the boundary conditions where $ r \rightarrow \infty $, then, the Abelian field $ C_{\mu} $ is a pure gauge on the boundary and does not contribute to the field strength tensor $\textbf{G}_{\mu\nu}$
of Eq. (\ref{15}). The energy density on the boundary must be zero; otherwise the string tension of the vortex will be infinite.
By this condition we can fix the value of $ a $ on the boundary.
For a static configuration, the energy density on the boundary $ H $ is
\begin{equation} 
H = - L = \frac{1}{4} \textbf{G}_{\mu\nu}. \textbf{G}^{\mu\nu} = \frac{1}{4} ( F_{\mu\nu} F^{\mu\nu} + B_{\mu\nu} B^{\mu\nu} +2 F_{\mu\nu} B^{\mu\nu}) \quad (r \rightarrow \infty),
\end{equation}
by choosing $ a = g $ one finds $ B_{\mu\nu} = 0 $ and $ H \rightarrow 0 $ as $ r \rightarrow \infty $, making possible a field configuration of finite energy. 

We use our decomposition, which is true for the infrared regime corresponding to 
the boundary of the system, in the above calculations. There must be a magnetic field parallel to the z axis at the center because 
the magnetic flux passing through the closed curve C at infinity is not zero.
Note that if our decomposition is valid for the boundary, then there is no singularity in the vortex core because the magnetic field can be finite.
The Abelian-Higgs model can be obtained \cite{Ahmad} if the 
general decomposition (\ref{8}) without the constraint (\ref{9}) is used .
Obviously, the Abelian-Higgs model supports vortex solutions known as Nielsen-Olesen vortices with finite string tension.

One can find all the homotopically inequivalent classes of the mapping (\ref{18}) and the corresponding
vortex configurations by the following replacement
\begin{eqnarray}
\varphi \rightarrow n \, \varphi. \label{25}
\end{eqnarray}
Then the scalar field $ \phi $ describes all homotopically inequivalent mappings of
 (\ref{18}) with the homotopy class $ Z $
\begin{equation}
Z = n \, \, \, \, \, \, (n \, \, \, \, \, \, integer), \label{26}
\end{equation}
corresponding to
\begin{equation}
C_{r} = C_{z} = 0 \, , \, \,\,\,\, C_{\varphi} = - \frac{n}{gr} \quad  (r \rightarrow \infty) , \label{27}
\end{equation}
for the vector potential. The magnetic flux is:
\begin{eqnarray}
\phi_{B} = - \frac{2\pi n}{g}. \label{29}
\end{eqnarray}

Finally, we should check if the solutions (\ref{19}) and (\ref{20}) at infinity when applied in $\textbf{G}^{\mu\nu}$ satisfy  the following equations
 of motion
\begin{eqnarray}
\textbf{n} . \triangledown_{\nu} \textbf{G}^{\mu\nu} = 0, \nonumber\\ &&
\hspace{-28mm} \partial_{\mu} \textbf{n} . \triangledown_{\nu} \textbf{G}^{\mu\nu} = 0, \nonumber\\ &&
\hspace{-28mm} (\textbf{n} \times \partial_{\mu} \textbf{n}) . \triangledown_{\nu} \textbf{G}^{\mu\nu} = 0, \nonumber\\ &&
\hspace{-28mm} (D_{\mu} \rho - D_{\mu} \sigma \textbf{n} \times) \triangledown_{\nu} \textbf{G}^{\mu\nu} = 0,
\end{eqnarray}
These equations are obtained from the Faddeev-Niemi Lagrangian \cite{Faddeev} which is valid for the whole space, not just the boundary. However, our decomposition
is valid for the boundary.
Since $ \textbf{G}^{\mu\nu} \rightarrow 0 $ as 
$ r \rightarrow\infty  $,
so our particular choices for $ \rho $, $ \sigma $ and $ C_{\mu} $ satisfy the above equations of motion.
This completes the vortex existence argument. 

In the next section we see that monopoles can also appear and vortices can confine them.

\section{\label{sec:level5}Monopoles in superconducting medium}

In addition to the vortices, Wu-Yang monopoles can also emerge in SU(2) Yang-Mills theory. According to Abelian dominance, the Abelian part of the SU(2) field strength tensor $ \textbf{G}_{\mu\nu} $ dominates in the infrared limit. So, we overlook the off-diagonal parts of $ \textbf{G}_{\mu\nu} $ to get
\begin{eqnarray}
\textbf{G}_{\mu\nu} = (F_{\mu\nu} + B_{\mu\nu}) \textbf{n} ,
\end{eqnarray}
where
\begin{eqnarray}
B_{\mu\nu} = \mu (\phi ^{\ast} \phi) H_{\mu\nu} ,
\end{eqnarray}
and
\begin{eqnarray}
\mu (\phi ^{\ast} \phi) = (1-\frac{\phi ^{\ast} \phi}{g^{2}}).
\end{eqnarray}
$ \mu (\phi ^{\ast} \phi) $ is a parameter characteristic of the medium; we call it ''vacuum permeability".
We have
\begin{eqnarray}
0 \leqslant \mu (\phi ^{\ast} \phi) \leqslant 1 .
\end{eqnarray}

We know that for SU(2) gauge theory, the magnetic charge which is a topological charge, can be described by the homotopy class
of the mapping $ \Pi_{2}(S^{2}) $ of the 2D sphere $ S^{2}_{R} $ to the coset space
$ S^{2} = SU(2) / U(1)$ of the internal space.
To obtain the magnetic field from $ H_{\mu\nu} $, we choose a hedgehog configuration for $ \textbf{n} $
\begin{equation}
\textbf{n}=\frac{r^{a}}{r}=
\begin{pmatrix}
\sin{\alpha} \, cos{\beta} \\
\sin{\alpha} \, sin{\beta} \\
\thickspace cos{\alpha}
\end{pmatrix}
\end{equation}
where
$$ \alpha = \theta, \quad
\beta = m \varphi. $$
$ \theta $ and $ \varphi $ are the angular spherical coordinates of $ S^{2}_{R} $, and $ m $
is an integer number. The magnetic intensity $ H $ can be obtained,
\begin{eqnarray}
\overrightarrow{H} = H \widehat{r}, \nonumber\\ &&
\hspace{-18mm} H = H_{\theta\varphi} = - \frac{1}{g} \textbf{n} . (\partial_{\theta} \textbf{n} \times \partial_{\varphi} \textbf{n}) 
 =  - \frac{m}{g} \frac{1}{r^{2}}, \label{31}
\end{eqnarray}
where
$$ \partial_{\theta} = \frac{\partial}{r\partial \theta} ,  \partial_{\varphi} = \frac{\partial}{r sin\theta\partial \varphi}. $$
From Eq. (\ref{31})
\begin{equation}
\overrightarrow{\nabla}.\overrightarrow{H} = - \frac{m}{g} 4\pi \delta(\overrightarrow{r}), \label{32}
\end{equation}
which means that there is a magnetic monopole in the origin.
Comparing Eq. (\ref{32}) with the Gauss equation for a magnetic monopole
charge, $ - g_{m} $:
$$ \overrightarrow{\nabla}.\overrightarrow{H} = - 4\pi g_{m} \delta(\overrightarrow{r}), $$
one gets
\begin{equation}
g  g_{m} = m , \label{33}
\end{equation}
which is the Dirac quantization condition.

We have
\begin{equation}
\overrightarrow{B} = \mu (\phi ^{\ast} \phi) \overrightarrow{H} .
\end{equation}
Note that the vacuum behaves like a superconducting medium in which the scalar field  $ \phi $ is a condensate. Therefore,
the ''vacuum permeability" depends on the value of the condensed field.
The magnetic field $ \overrightarrow{B} $ depends on the ''vacuum permeability" and in the Higgs vacuum where $ \phi ^{\ast} \phi = g^{2} $, it goes to zero. 
So the vacuum that is structured by the Higgs field $ \phi $ does not allow the presence of the magnetic field except in the flux-tube (vortex) form. These vortices can confine monopoles. We recall that the existence of the monopoles has been studied in Cho 
decomposition \cite{Cho}. But we have discussed the effect of the condensed field, $\phi$, on the monopoles.

\section{\label{sec:level6}Summary and discussion}

We have looked for a suitable parameterization of the SU(2) Yang-Mills field in the low-energy limit that helps to uncover the vacuum structure,  particularly topological objects, like 
vortices and monopoles, believed in some models to be related to the phenomenon of confinement.
We conjecture that for the low-energy limit of the SU(2) Yang-Mills theory, the field strength
tensor can be obtained by 
multiplying two parts, $ \textbf{G}_{\mu\nu}=G_{\mu\nu} \textbf{n} $. The first part, $ G_{\mu\nu} $,  
is a tensor with the space-time indices and the second part $ \textbf{n} $, is a  three-components unit vector field that selects the Abelian direction at each
space-time point. This conjecture is motivated by the form of
the field strength tensor of Cho's restricted theory. We propose a decomposition of the SU(2) Yang-Mills field corresponding
to this idea. It is similar to the Faddeev-Niemi decomposition, but with the constraint $ D_{\mu} \phi = (\partial_{\mu} + igC_{\mu}) \phi = 0 $.
This constraint leads to the appearance of vortices with finite energy per unit length.

The constraint $ D_{\mu} \phi = (\partial_{\mu} + igC_{\mu}) \phi = 0 $ is similar to the constraint of Cho's restricted theory $ D_{\mu} \textbf{n} = (\partial_{\mu} + g \textbf{A}_{\mu}\times) \textbf{n} = 0 $. If one generalizes the condition $ D_{\mu} \phi = 0 $ to be valid not only in the infrared regime but also for the whole energy spectrum, then 
one gets string-like singularities \cite{Mo}. Therefore,
it seems that vanishing  the covariant derivative of the scalar field in
an Abelian or non-Abelian Yang-Mills theory has something to do with the low-energy limit and the vacuum structure of the theory.

Besides  vortices, Wu-Yang monopoles can also appear by choosing a hedgehog ansatz for $ \textbf{n} $. However, they appear in a medium that behaves like a superconductor 
with "vacuum permeability" $ \mu $ which depends on the value of the condensate field $ \phi $. The magnetic field goes to zero in the Higgs vacuum where $ \mu = 0 $.
Therefore, magnetic fields cannot penetrate the vacuum which behaves like a superconductor medium except in a flux-tube shape. Hence, magnetic monopoles will be confined.

\section{\boldmath Acknowledgments}
We are grateful to the Research Council of the University of Tehran for supporting this study.


\begin{thebibliography}{99}

\bibitem{'Hooft}
G. 't Hooft, Nucl. Phys. B \textbf{153}, 141 (1979);
A. Polyakov, Nucl. Phys. B \textbf{120}, 429 (1977).

\bibitem{Nambu}
Y. Nambu, Phys. Rev. D \textbf{10} 4262 (1974);
M. Creutz, Phys. Rev. D \textbf{10} 2696 (1974);
G. 't Hooft, High Energy Physics, Editorice Compositori, Bologna (1975);
S. Mandelstam, Phys. Report \textbf{23} 245 (1976).

\bibitem{Ripka}
G. Ripka, Dual Superconductor Models of Color Confinement, Springer-Verlag, 2005, arXiv: 0806.1078v2 [hep-th].

\bibitem{Abelian}
G. 't Hooft, Nucl. Phys. B \textbf{190} 455 (1981).

\bibitem{Cho}
Y. M. Cho, Phys. Rev. D \textbf{21} (1980) 1080; D \textbf{23} 2415 (1981).

\bibitem{Faddeev}
L. Faddeev and A. J. Niemi, Phys. Rev. Lett. \textbf{82} 1624 (1999).

\bibitem{Shabanov}
S. V. Shabanov, Phys. Lett. B \textbf{458} 322 (1999); Phys. Lett. B \textbf{463} 263 (1999).

\bibitem{Yotsuyanagi}
T. Suzuki and I. Yotsuyanagi, Phys. Rev. D \textbf{42} 4257 (1990).

\bibitem{Stack}
J. D. Stack, S. D. Neiman, and R. Wensley, Phys. Rev. D \textbf{50} 3399 (1994);
H. Shiba and T. Suzuki, Phys. Lett. B \textbf{333} 461 (1994).

\bibitem{Suganuma}
T. Suzuki, Prog. Theor. Phys. \textbf{80} 929 (1988);
H. Suganuma, S. Sasaki, and H. Toki, Nucl. Phys. B \textbf{435} 207 (1995).

\bibitem{Deldar}
L. S. Grigorio, M. S. Guimaraes, W. Oliveira, R. Rougemont, and C. Wotzasek,
Phys. Lett. B \textbf{697} 392 (2011);
S. Deldar and A. Mohamadnejad, Phys. Rev. D \textbf{86} 065005 (2012).

\bibitem{Wu}
T. T. Wu and C. N. Yang, in Properties of Matter Under Unusual Conditions, edited by
H. Mark and S. Fernbach (Interscience, New York, 1969).

\bibitem{Oxman}
L. E. Oxman, JHEP \textbf{12} 89 (2008).

\bibitem{Evslin}
J. Evslin and S. Giacomelli, JHEP \textbf{4} 22 (2011).

\bibitem{Niemi3}
A. J. Niemi and A. Wereszczynski, J. Math. Phys. \textbf{52} 072302 (2011).

\bibitem{Pak}
Y. M. Cho, H. W. Lee, and D. G. Pak, Phys. Lett. B \textbf{525} 347 (2002).

\bibitem{Bae}
W. S. Bae, Y. M. Cho, and S. W. Kimm, Phys. Rev. D \textbf{65} 025005 (2001).

\bibitem{Niemi1}
L. Faddeev and A. J. Niemi, Phys. Lett. B \textbf{464} 90 (1999).

\bibitem{Niemi2}
L. Faddeev and A. J. Niemi, Phys. Lett. B \textbf{525} 195 (2002).

\bibitem{Mo}
S. Deldar and A. Mohamadnejad, PoS ConfinementX (2012) 290; arXiv: 1301.2057v1 [hep-th].

\bibitem{Ahmad}
A. Mohamadnejad and S. Deldar, Mod. Phys. Lett. A \textbf{29} 1450047 (2014).

\end{thebibliography}
\end{document}